\documentclass[a4paper,pdftex,floatfix,groupedaddress,pra,showpacs,twocolumn]{revtex4}
\pdfoutput=1

\usepackage{amsfonts,amsmath,amssymb}
\usepackage{graphicx}
\usepackage{hypernat}
\usepackage[utf8]{inputenc}
\usepackage[hyperfootnotes=false]{hyperref}
\usepackage{textcomp}
\hypersetup{colorlinks,citecolor=black,urlcolor=blue}

\newlength{\figwidth}
\newlength{\figwidthsmall}
\newcommand{\degree}{\ensuremath{^\circ}}%
\newcommand{\ie}{i.\,e.}%
\newcommand{\cost}{\ensuremath{\langle\cos^2\theta_{2D}\rangle}}

\begin{document}

\title{Quantum-state selection, alignment, and orientation of large molecules using static electric
   and laser fields}%
\author{Frank Filsinger}%
\author{Jochen K\"{u}pper}%
\email{jochen@fhi-berlin.mpg.de}%
\author{Gerard Meijer}%
\affiliation{Fritz-Haber-Institut der Max-Planck-Gesellschaft, Faradayweg 4-6, 14195 Berlin,
   Germany}

\author{Lotte Holmegaard$^1$}%
\author{Jens H. Nielsen$^2$}%
\author{Iftach Nevo$^{1}$}%
\author{Jonas L. Hansen$^{3}$}%
\author{Henrik Stapelfeldt$^{1,3}$}%
\email{henriks@chem.au.dk}%
\affiliation{$^1$Department of Chemistry, University of Aarhus, 8000 Aarhus C, Denmark \\
   $^2$Department of Physics and Astronomy, University of Aarhus, 8000 Aarhus C, Denmark \\
   $^3$Interdisciplinary Nanoscience Center (iNANO), University of Aarhus, 8000 Aarhus C, Denmark}%

\date{\today}%
\begin{abstract}\noindent%
   Supersonic beams of polar molecules are deflected using inhomogeneous electric fields. The
   quantum-state selectivity of the deflection is used to spatially separate molecules according to
   their quantum state. A detailed analysis of the deflection and the obtained quantum-state
   selection is presented. The rotational temperatures of the molecular beams are determined from
   the spatial beam profiles and are all approximately 1~K.
   Unprecedented degrees of laser-induced alignment $(\cost=0.972)$ and orientation of iodobenzene
   molecules are demonstrated when the state-selected samples are used. Such state-selected and oriented
   molecules provide unique possibilities for many novel experiments in chemistry and physics.
\end{abstract}
\maketitle%

\section{Introduction}

For a large range of experiments in chemistry and physics, a high level of control over the external
and internal degrees of freedom of molecules is very beneficial. This includes control over
the translational and the rotational motions, as well as the selection of a single quantum state or
a small set of states. Such quantum-state-selected targets provide unique possibilities, for
example, for manipulating the external degrees of freedom with static electric
fields~\cite{Loesch:JCP93:4779, Friedrich:Nature353:412} or optical
fields~\cite{Friedrich:PRL74:4623, Stapelfeldt:RMP75:543}, or both~\cite{Friedrich:JCP111:6157,
   Friedrich:JPCA103:10280}. The quantum-state selection also naturally discriminates between
individual stereo-isomers of large molecules~\cite{Filsinger:PRL100:133003}. The resulting samples
of aligned or oriented individual isomers offer unique prospects for novel experiments with complex
molecules, such as femtosecond pump-probe measurements, x-ray or electron diffraction in the
gas-phase~\cite {Spence:PRL92:198102, Peterson:APL92:094106}, high-harmonic
generation~\cite{Levesque:PRL99:243001}, or tomographic reconstructions of molecular
orbitals~\cite{Itatani:Nature432:867}. Moreover, it would provide considerably increased control in
reaction dynamics experiments~\cite{Stolte:StateSelectedScattering:1988}.

Strong cooling can be achieved in supersonic expansions of molecules seeded in an inert atomic
carrier gas. For small molecules (consisting of just a few atoms) only a few rotational states are
populated at the typical temperatures on the order of 1~K. For larger polyatomic systems rotational
cooling down to or even below 1 K still leaves the molecular ensemble distributed over a
considerable number of rotational states, thereby often masking quantum-state-specific effects.
State selection can be performed using inhomogeneous electric or magnetic fields. The possibility to
deflect polar molecules in a molecular beam with an electric field was first described by Kallmann
and Reiche in 1921~\cite{Kallmann:ZP6:352} and experimentally demonstrated by Wrede in
1927~\cite{Wrede:ZP44:261}. As early as 1926, Stern suggested that the technique could be used for
the quantum-state separation of small diatomic molecules at low temperatures~\cite{Stern:ZP39:751}.
In 1939 Rabi introduced the molecular beam resonance method, by using two deflection elements of
oppositely directed gradients in succession, to study the quantum structure of atoms and
molecules~\cite{Rabi:PR55:526}. Whereas deflection experiments allow the spatial dispersion of
quantum states, they do not provide any focusing. For small molecules in low-field-seeking states
this issue could be resolved using multipole focusers with static electric fields. These were
developed independently in 1954/55 in Bonn~\cite{Bennewitz:ZP139:489, Bennewitz:ZP141:6} and in New
York, where they were used to produce the population inversion for the first MASER
experiments~\cite{Gordon:PR95:282, Gordon:PR99:1264}. About ten years later, molecular samples in a
single rotational state were used for state specific inelastic scattering experiments by the Bonn
group~\cite{Bennewitz:ZP177:84} and, shortly thereafter, for reactive
scattering~\cite{Brooks:JCP45:3449, Beuhler:JACS88:5331}. In the following decades, multipole
focusers were extensively used to study steric effects in gas-phase reactive scattering
experiments~\cite {Parker:ARPC40:561, Stolte:StateSelectedScattering:1988}. Also for the
investigation of steric effects in gas-surface scattering~\cite{Kuipers:Nature334:420} and
photodissociation~\cite{Rakitzis:Science303:1852} experiments, the preparation of oriented samples
of state-selected molecules using electrostatic focusers was essential. For about ten years, it is
also possible to manipulate the speed of small molecules using switched inhomogeneous electric
fields in the so-called Stark decelerator~\cite{Bethlem:PRL83:1558}. More recently, also its
optical~\cite{Fulton:PRL93:243004} and magnetic~\cite{Vanhaecke:PRA75:031402} analogs have been
demonstrated.

Obtaining similar control over large molecules is more difficult, because all low-lying quantum
states are high-field seeking at the required electric field strengths. In order to confine these
molecules, dynamic focusing schemes are necessary~\cite{Auerbach:JCP45:2160, Bethlem:JPB39:R263,
   Tarbutt:NJP10:073011}. Dynamic focusing of large molecules has been demonstrated in the
alternating-gradient (AG) deceleration of benzonitrile~\cite{Wohlfart:PRA77:031404} and in the
conformer selection of 3-aminophenol~\cite{Filsinger:PRL100:133003}. However, if focusing is not
necessary, spatial dispersion of quantum states can still be achieved using static electric fields
in a Stern-Gerlach-type deflector. This molecular beam deflection has been used extensively as a
tool to determine dipole moments and polarizabilities of molecular systems ranging from
diatomics~\cite{Wrede:ZP44:261} over clusters~\cite{Moro:Science300:1265} to large
biomolecules~\cite{Broyer:PhysScr76:C135}.

Recently, we have demonstrated the quantum-state selection of large
molecules~\cite{Holmegaard:PRL102:023001} following the original proposal of
Stern~\cite{Stern:ZP39:751}. Here, the details of the electrostatic deflection are presented. It is
shown, how the rotational temperature of cold supersonic jets can be determined with high precision
from deflection measurements and that indeed a small subset of quantum states can be addressed in
deflected samples of large molecules. In particular, the ground state has the largest Stark shift
and molecules residing in this state are deflected the most. Our goal is to isolate and use
rotational ground state molecules, or at least samples of molecules in the few lowest lying states,
as targets for various experiments. Since the deflection does not change the initial state
distribution but merely disperses it, it is crucial that the population of ground state molecules in
the molecular beam is initially as large as possible. Therefore, the rotational temperature of the
molecular beam is made as low as possible using a high-pressure supersonic expansion~\cite{Hillenkamp:JCP118:8699}. It is shown how the resulting state-selected molecules can be used
to improve one-dimensional (1D) laser-induced alignment~\cite{Stapelfeldt:RMP75:543,
   Kumarappan:JCP125:194309} and mixed field orientation~\cite{Buck:IRPC25:583,
   Minemoto:JCP118:4052, Holmegaard:PRL102:023001}. Here, alignment refers to confinement of a
molecule-fixed axis (typically, the largest polarizability axis) along a laboratory-fixed axis and
orientation refers to the molecular dipole moments pointing in a particular direction. Alignment and
orientation occur in the adiabatic limit where the laser field, used to align the molecules, is
turned on and off slowly compared to the inherent rotational periods of the molecule
\cite{Friedrich:JCP99:15686, Sakai:JCP110:10235}. The state selection leads to strong enhancement in
the degree of orientation and alignment of iodobenzene molecules compared to that achieved when no
deflection is used.

\section{Experimental}
\label{sec:experimental}

\begin{figure}
   \centering
   \includegraphics[width=\figwidth]{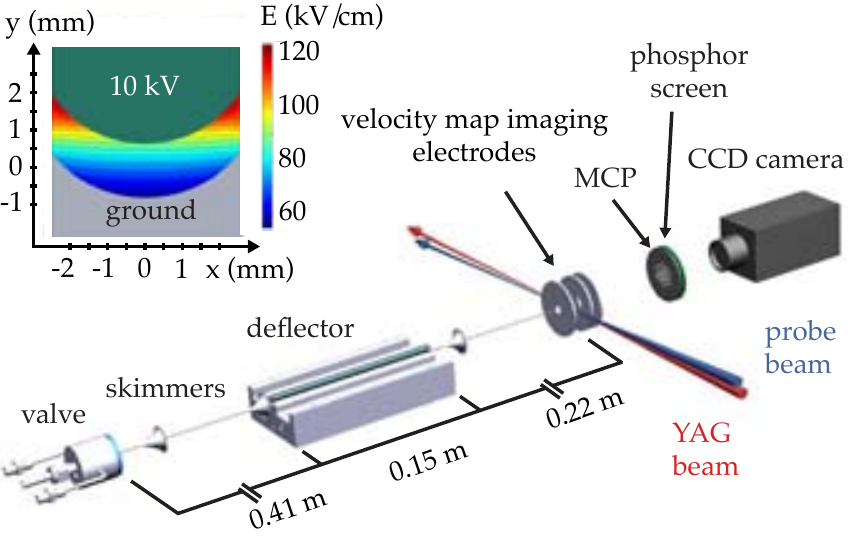}
   \caption{(Color online): Scheme of the experimental setup. In the inset, a cut through the
      deflector is shown, and a contour-plot of the electric field strength is given. Details of the
      velocity map imaging spectrometer are shown in \autoref{fig:results:alignment:probegeometry}.
      See text for details. }
   \label{fig:experimental:setup}
\end{figure}

A schematic of the experimental setup is shown in \autoref{fig:experimental:setup}.
The molecular beam machine consists of three differentially pumped vacuum chambers; The source
chamber housing a pulsed valve (pumped by a 2000~l/s turbomolecular pump), the deflector chamber
(pumped by a 500~l/s turbomolecular pump) and the detection chamber housing the ion/electron
spectrometer (pumped by a 500~l/s turbomolecular pump). About 3~mbar of iodobenzene (Sigma Aldrich,
98~\% purity) or benzonitrile (Sigma Aldrich, 98~\% purity) is seeded in an inert carrier gas and
expanded through a pulsed valve into vacuum. In order to obtain optimal cooling of the molecular
beam, a miniaturized, high pressure Even-Lavie valve~\cite{Hillenkamp:JCP118:8699} is used operating
at a backing pressure of 90~bar of He or 20~bar of Ne, limited by the onset of cluster formation.
While rotational temperatures down to 0.4~K have been achieved under similar
conditions~\cite{Even:JCP112:8068}, the typical rotational temperature in our experiments is
$\sim1$~K. Two 1-mm-diameter skimmers placed 15~cm (separating the source and the
deflector chamber) and 38~cm downstream from the nozzle collimate the molecular beam before it
enters a 15-cm-long electrostatic deflector. A cut through the electrodes of the deflector is shown
in the inset of \autoref{fig:experimental:setup} together with the electric field created. A trough
with an inner radius of curvature of 3.2~mm at ground potential and a rod with a radius of 3.0~mm at
high voltage create a two-wire field~\cite{Ramsey:MolBeam:1956}. The vertical gap across the
molecular beam axis is 1.4~mm, while the smallest distance between the electrodes is 0.9~mm. The
two-wire field geometry is ideally suited for molecular beam deflection. The gradient of the
electric field along the vertical direction is large and nearly constant over a large area explored
by the molecular beam, while the electric field is very homogeneous along the horizontal direction.
Thus, a polar molecule experiences a nearly constant force in the vertical direction independent of
its position within the deflector, while the force in the horizontal direction (\ie, broadening of
the beam in the horizontal direction) is minimized. In our setup, the deflector is mounted such that
molecules in high-field-seeking (low-field-seeking) quantum states are deflected upwards
(downwards).

After passing through the deflector, the molecular beam enters the differentially pumped detection
chamber via a third skimmer of 1.5~mm diameter. In the detection area, the molecular beam is crossed
by one or two laser beams that are focused by a spherical lens with a focal length of $f=300$~mm.
The lens is mounted on a vertical translation stage so that the height of the laser foci can be
adjusted with high precision. In the first part of the experiment, where the beam deflection of
iodobenzene and benzonitrile is characterized, only one laser, the probe laser, is used. This
Ti:Sapphire laser (25~fs (FWHM) pulses, 800~nm, beam-waist $\omega_0=21$~\textmu{}m) is used to
determine the relative density in the molecular beam via photoionization. In the second part of the
experiment, an additional laser pulse is included to study laser-induced alignment and orientation
of iodobenzene. For these experiments, 10~ns (FWHM) long pulses from a Nd:YAG laser (1064~nm,
$\omega_0=36$~\textmu{}m) are overlapped in time and space with the probe laser pulses. While the
YAG laser induces adiabatic alignment and orientation, here the fs-laser is used to determine the
spatial orientation of the target molecules via Coulomb explosion. Ionic fragments produced in the
Coulomb explosion are accelerated in a velocity focusing geometry towards the detector. The detector
can be gated with a time resolution of $\sim90$~ns, which allows for mass selective detection of
individual fragments. A microchannel plate (MCP) detector backed by a phosphor screen is employed to
detect the position of mass-selected ions. In particular, I$^+$ fragment ions, formed in the Coulomb
explosion of iodobenzene, are particularly useful experimental observables since they recoil along
the C-I symmetry axis of the molecule. Thus, 2D ion images of I$^+$ recorded with a CCD camera
provide direct information about the instantaneous molecular orientation of the C-I bond axis with
respect to the laboratory frame and are, therefore, the basic observables in these experiments. All
experiments are conducted at 20~Hz, limited by the repetition rate of the YAG laser.

\section{Results and Discussion}
\label{sec:results}

In the first experiments shown in \autoref{ssec:results:deflection}, a detailed analysis of
electrostatic beam deflection is presented. It is shown, that electrostatic deflection of cold
molecular beams can be used to determine the rotational temperature of a supersonic jet.
Furthermore, the degree of quantum-state selectivity that can be achieved with our setup is
investigated. As an application it is demonstrated, how this quantum-state selectivity of the
deflection process can be exploited to obtain an unprecedented degree of laser-induced alignment
(\autoref{ssec:results:alignment}) and orientation (\autoref{ssec:results:orientation}) of
iodobenzene molecules.

\subsection{Electrostatic Deflection of Cold Molecular Beams}
\label{ssec:results:deflection}

In the first experiment, the deflection of benzonitrile molecules (BN, C$_7$H$_5$N) seeded in 90~bar
of He is investigated. BN is an ideal candidate for electrostatic beam deflection due to its large
permanent dipole moment of 4.515~D. From the precisely known molecular
constants~\cite{Wohlfart:JMolSpec247:119} the energy of a given rotational quantum state
can be calculated as a function of the electric field strength. The exact procedure is detailed in
\autoref{appendix:stark}. \autoref{fig:results:starkBN} shows the Stark energies for the lowest
rotational states of BN.
\begin{figure}
   \centering
   \includegraphics[width=\figwidthsmall]{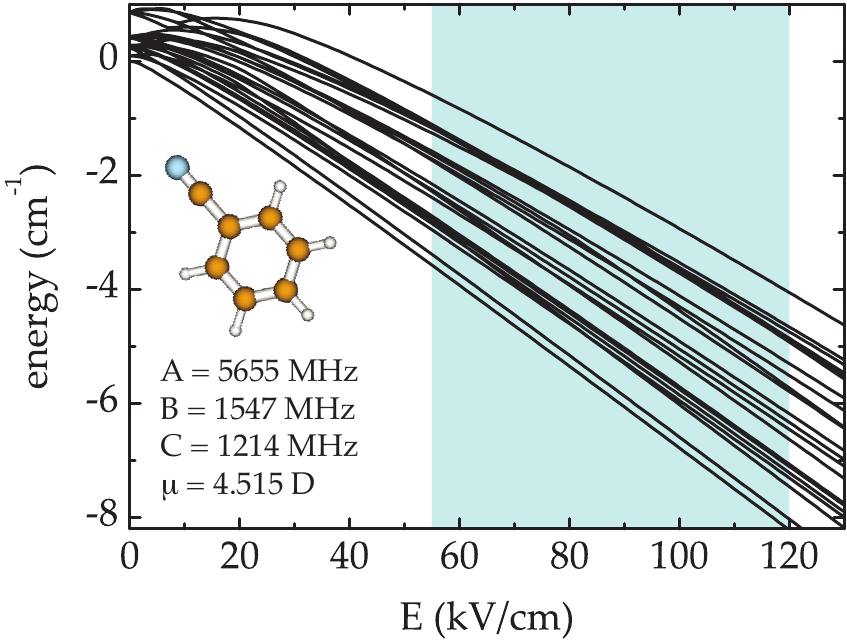}
   \caption{(Color online): Energy as a function of the electric field strength for the
      lowest rotational quantum states of benzonitrile. The shaded area indicates the actual
      electric field strength inside the deflector. In the inset, the molecular structure is shown
      together with the relevant molecular constants~\cite{Wohlfart:JMolSpec247:119}.}
   \label{fig:results:starkBN}
\end{figure}
Due to the small rotational constants and the resulting high density of rotational states, a large
number of states is populated even under the cold conditions in a supersonic expansion. At a
rotational temperature of 1~K, the typical temperature in our experiments (\emph{vide infra}), 66
rotational quantum states (with 419 $M$-components) have a population larger than 1~\%
relative to the ground state. At the electric field strengths present in the deflector, indicated by
the shaded area in \autoref{fig:results:starkBN}, all low-lying quantum states are high-field
seeking. This is due to mixing of closely spaced states of the same symmetry and is typical for large
asymmetric top molecules. The Stark shift and thus the force a molecule experiences in an
inhomogeneous electric field depends on the rotational quantum state. Molecules in the ground state
have the largest Stark shift and are, therefore, deflected the most. In general, the Stark shift
decreases with increasing $J$ quantum number. Thus, the lower the rotational temperature of the
molecular beam is, the more the beam is deflected.

\begin{figure}
   \centering
   \includegraphics[width=\figwidth]{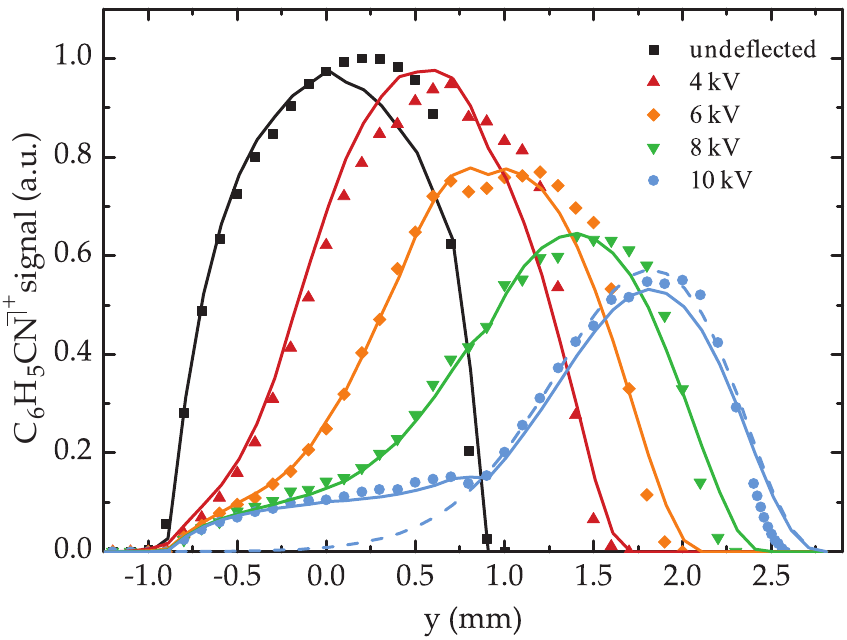}
   \caption{(Color online): The vertical spatial profile of the molecular beam for different
      deflection voltages applied, measured by recording the laser-induced BN$^+$ signal (see text).
      The experimental data are shown as symbols together with the corresponding simulated profiles (lines). }
   \label{fig:results:deflection:BN}
\end{figure}

\autoref{fig:results:deflection:BN} shows vertical intensity profiles of BN for various high
voltages applied to the deflector.
Vertical intensity profiles are obtained by recording the BN$^+$ signal from photoionization by the
fs-laser as a function of the vertical position of the laser focus. If no high-voltage is applied to
the deflector, the molecular beam extends over about $\sim2$~mm. In this case, the size of the
molecular beam in the detection region is determined by the mechanical aperture of the experimental
setup, \ie, by the dimensions of the deflector and the last skimmer before the detection region. As
the high voltage is turned on, the molecular beam profile broadens and shifts upwards. At a voltage
of 10~kV, a large fraction of the molecules is deflected out of the original, undeflected beam
profile. A small fraction of the molecules, however, is almost unaffected by the deflector.

In order to understand these experimental findings, Monte Carlo simulations are employed, which are
described in detail in~\autoref{appendix:simulations}. In brief, trajectory calculations are
performed for molecular packets of individual rotational quantum states. These calculations yield
single-quantum-state deflection profiles. Then, the single-state profiles are averaged according to
the populations of the respective states in the original molecular beam (\ie, at the entrance of the deflector). From these simulations it
is obvious, that the molecules in the original beam are not rotationally thermalized, an effect that
has previously been observed in rotationally resolved spectroscopy~\cite{Wu:JCP91:5278,
   Berden:JCP103:9596}. A number of different descriptions of the populations of rotational states
have been given~\cite{Berden:JCP103:9596}; we use the formula for a two-temperature model originally
proposed by Levy and coworkers~\cite{Wu:JCP91:5278}. For details and the approximation of the
high-temperature component see~\autoref{appendix:profiles}. Finally, the rotational temperature of
the low-temperature component in the molecular beam is obtained by fitting the simulated deflection
profiles to the experimental data using a local optimization algorithm. All deflection profiles
measured at the different voltages are fitted simultaneously, where the fraction $q$ of the
low-temperature component, a general intensity scaling factor $s$ of the deflected profiles (with
respect to the undeflected beam profile), and the rotational temperature $T_{\text{rot}}$ of the
low-temperature component are the fitting parameters. Best agreement between experimental data and
simulations is found for $q=0.93$ and $T_{\text{rot}}=0.8$~K. The resulting simulated deflection
profiles nicely reproduce the experimental data as shown in
\autoref{fig:results:deflection:BN}~(solid lines). In particular, the undeflected part of the
molecular beam for 10~kV can be perfectly simulated, which indicates that the use of a
two-temperature model was indeed justified. For comparison, also a simulated deflection profile for
10~kV using a one-temperature model is shown (dashed line in \autoref{fig:results:deflection:BN}).

\begin{figure}
   \centering
   \includegraphics[width=\figwidth]{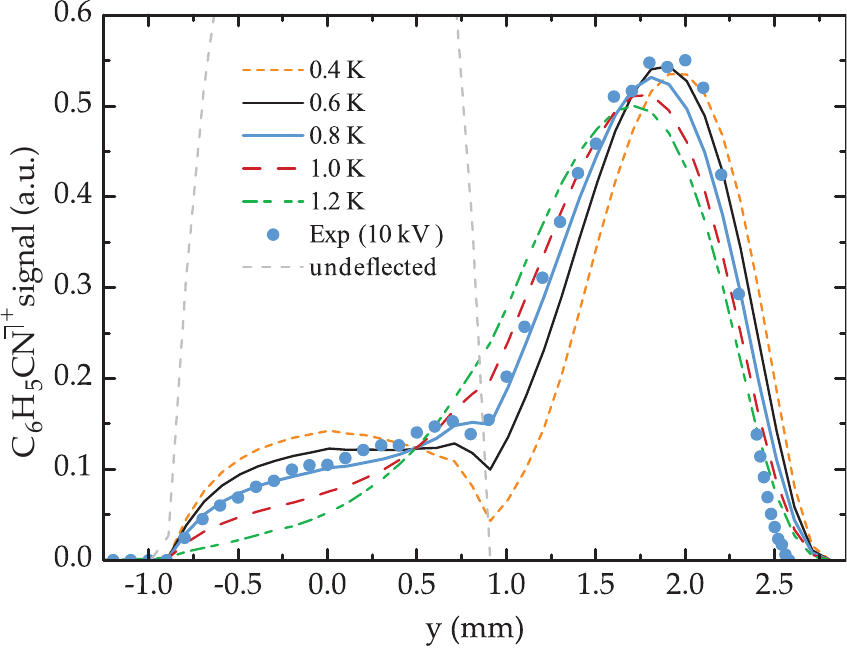}
   \caption{(Color online): The vertical profile of the molecular beam measured by recording the
      laser-induced BN$^+$ signal (see text). The experimental data are shown together with the
      corresponding simulated profiles (solid lines) for different rotational temperatures.
      Simulations are shown for the two-temperature model; see text for details.}
   \label{fig:results:deflection:BN_Temp}
\end{figure}

In order to estimate the uncertainty of $T_{\text{rot}}$, deflection profiles are calculated for
different rotational temperatures. For each fixed rotational temperature, the best values for $s$
and $q$ are determined using the fitting procedure outlined above and the resulting deflection
profile for a voltage of 10~kV is plotted in \autoref{fig:results:deflection:BN_Temp}.
With increasing $T_{\text{rot}}$, the peak of the beam profile shifts towards smaller y-values,
while, at the same time, the intensity in the undeflected part of the beam profile is reduced. From
the comparison of experimental data and simulation, an uncertainty of $T_{\text{rot}}$ of $\pm0.2$~K
is estimated.

\begin{figure}
   \centering
   \includegraphics[width=\figwidth]{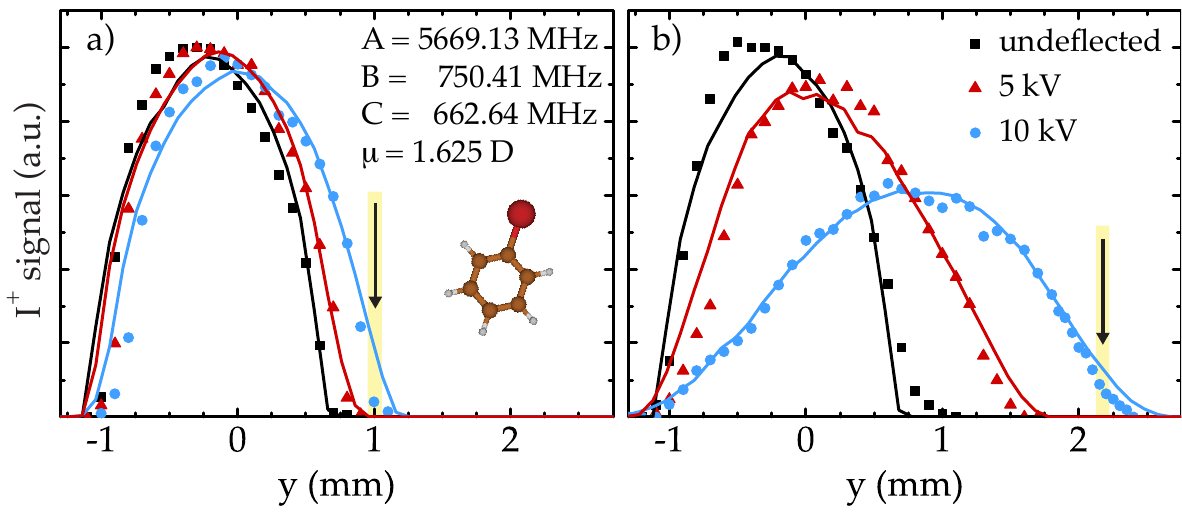}
   \caption{(Color online): The vertical profile of the molecular beam for different deflection
      voltages applied, measured by recording the laser-induced I$^+$ signal (see text). The
      experimental data are shown together with the corresponding simulated profiles. Figure (a)
      shows the beam profiles for iodobenzene seeded in 90 bar He, and Figure (b) shows the
      respective profiles for iodobenzene seeded in 20 bar Ne. Arrows indicate the laser positions
      for alignment and orientation experiments (\emph{vide infra}).}
   \label{fig:results:deflection:IB}
\end{figure}

The deflection of iodobenzene molecules (IB, C$_6$H$_5$I) is investigated in the same way. Vertical
intensity profiles are measured by recording the signal of I$^+$ ions, created by Coulomb explosion
with a circularly polarized probe pulse, as a function of the vertical position of the probe laser
focus. \autoref{fig:results:deflection:IB}~a) shows deflection measurements for IB seeded in 90~bar
of He.
IB (mass 204~u) is heavier than BN (mass 103~u) and has a considerably smaller dipole moment of only
1.625~D~\cite{Dorosh:JMolSpec246:228} compared to BN. Therefore, smaller deflection amplitudes are
observed for IB under identical expansion conditions. However, the interaction time with the
electric field and the time-of-flight from deflector to detection region can be increased when Ne is
used as a carrier gas instead of He. Changing the carrier gas reduces the mean velocity of the
molecular beam from $\sim1800$~m/s to $\sim800$~m/s and significantly enhances the observed
deflection as shown in \autoref{fig:results:deflection:IB}~b). Following the fitting procedure
outlined above, $T_{\text{rot}}$ can be determined for IB as well. In the case of IB, a
one-temperature model with a rotational temperature of 1.05~K fits the experimental data best for
the deflection measurements in helium as well as in neon. The uncertainty of $T_{\text{rot}}$ is
estimated to be $\pm0.1$~K for IB seeded in Ne and $\pm0.2$~K for IB in He. The somewhat larger
uncertainty for the measurements in He reflects the small deflections observed for He. The simulated
deflection profiles for IB are shown as solid lines in \autoref{fig:results:deflection:IB} and agree
well with the experimental data.

\begin{figure}
   \centering
   \includegraphics[width=\figwidth]{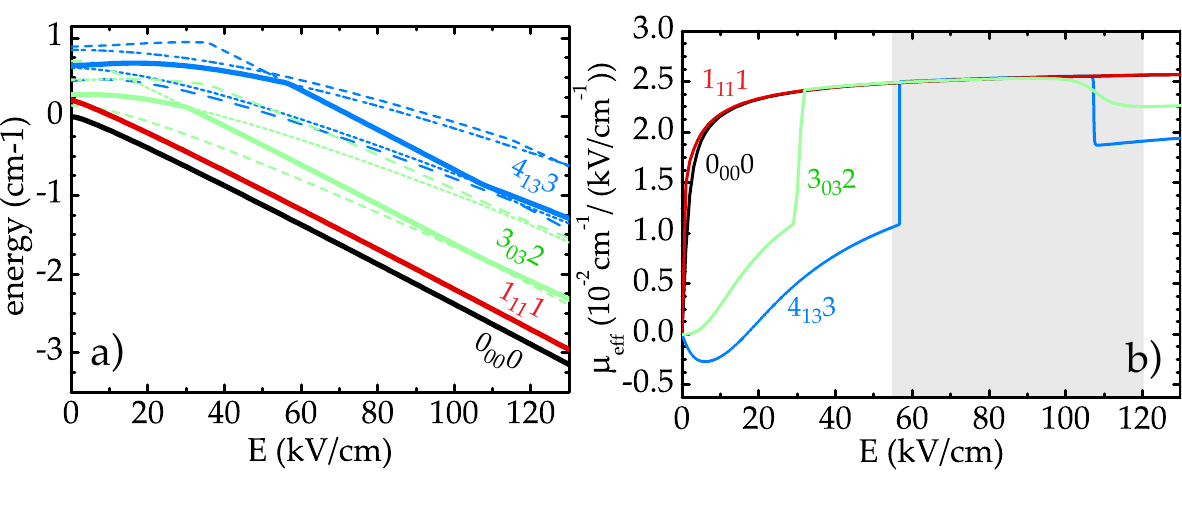}
   \caption{(Color online): (a) Energy as a function of the electric field strength for
      selected quantum states of IB. Solid lines represent quantum states that are present close to
      the upper cut-off of the molecular beam profile for IB seeded in Ne. The respective
      $J_{K_aK_c}$ quantum numbers are given in the figure. (b) Effective dipole moment for selected
      quantum states of IB. The shaded area represents the range of electric field strengths in the
      deflector at 10~kV. }
   \label{fig:results:deflection:starkIB}
\end{figure}

The main purpose of the deflection studies presented in this work is to provide
quantum-state-selected samples of large molecules for further experiments. The degree of deflection
that a molecule experiences in the electric field of the deflector depends on its quantum state. The
relevant quantity is the effective dipole moment $\mu_{\text{eff}}$ (the negative slope of the Stark
curve), which depends on the electric field strength. Molecules residing in low rotational quantum
states have generally the largest $\mu_{\text{eff}}$ and are, therefore, deflected most. These
molecules can simply be addressed by moving the laser focus in the detection region towards the
upper cut-off of the molecular beam profile. In order to understand the laser-induced alignment and
orientation experiments presented in \autoref{ssec:results:alignment} and
\ref{ssec:results:orientation}, it is crucial to know the relative populations of individual quantum
states that are probed at a given height of the laser focus. The positions of the laser foci within
the molecular beam profile during the alignment and orientation measurements are indicated by arrows
in \autoref{fig:results:deflection:IB}. At this position, the intensity of the deflected beam is
9~\% of the undeflected peak intensity. The composition of the molecular packets in the detection
region can be extracted from the simulated deflection profiles. \autoref{tab:results:deflection:pop}
provides an overview of the most abundant quantum states present in different regions of the beam
profiles for IB at a rotational temperature of 1.05~K.
\begin{table*}
\centering\small
\begin{tabular}{cc|cc|cc|cc|c}
   &  & \multicolumn{2}{c|}{IB in He at $0.01 \cdot I_{\text{peak}}$} & \multicolumn{2}{c|}{IB in Ne at $0.01
      \cdot I_{\text{peak}}$}& \multicolumn{2}{c|}{IB in Ne at $0.09 \cdot I_{\text{peak}}$} & undeflected beam\\
   $J_{K_aK_c}$&$M$ &$P_M$(\%)&$\Sigma P_M$(\%)&$P_M$(\%)&$\Sigma P_M$(\%)&$P_M$(\%)&$
   \Sigma P_M$(\%)&$P_{J_{K_aK_c}}^{\text{free}}$(\%)\\
   \hline
   \hline
   $0_{0  0}$    & 0 & 17.51 & 17.51 & 26.06 & 26.06 & 6.24 & 6.24 & 1.15 \\
   \hline
   $1_{0  1}$    & $ \begin{array}{cc} 0 \\ 1\\ \end{array}$ & $ \left. \begin{array}{cc} 0.02 \\ 7.69 \end{array}
   \right\} $ & 7.71 & & & $ \left. \begin{array}{cc} 3.90 \\ 9.70 \end{array} \right\} $ & 13.61 & 3.24\\
   \hline
   $1_{1  1}$    & $ \begin{array}{cc} 0 \\ 1\\ \end{array}$ & $ \left. \begin{array}{cc} 1.92 \\ 15.66 \end{array}
   \right\} $ & 17.58 & $ \left. \begin{array}{cc}  \\ 23.84 \end{array} \right\} $ & 23.84 & $ \left. \begin{array}
         {cc} 2.37 \\ 5.43 \end{array} \right\} $ & 7.79 & 1.55\\
   \hline
   $1_{1  0}$    & $ \begin{array}{cc} 0 \\ 1\\ \end{array}$ & $ \left. \begin{array}{cc} 1.81 \\ 0.02 \end{array}
   \right\} $ & 1.82 &  &  & $ \left. \begin{array}{cc} 2.41 \\ 3.86 \end{array} \right\} $ & 6.26 & 1.55\\
   \hline
   $2_{0  2}$    & $ \begin{array}{cc} 1 \\ 2\\ \end{array}$ &  &   &  &  & $ \left. \begin{array}{cc} 0.57 \\ 7.17 \end
      {array} \right\} $ & 7.74 & 4.75\\
   \hline
   $2_{1  2}$    & $ \begin{array}{ccc} 0\\1\\ 2\\ \end{array}$ & $ \left. \begin{array}{ccc} \\ \\3.22 \end{array}
   \right\} $  & 3.22  &  &  & $ \left. \begin{array}{ccc} 0.23 \\ 3.51\\4.16 \end{array} \right\} $ & 7.90 & 2.28\\
   \hline
   $3_{0  3}$    & $ \begin{array}{cccc} 0\\1\\ 2\\ 3\\ \end{array}$ & $ \left. \begin{array}{cccc} 0.01\\6.17\\23.55\\ \, \\ \end{array} \right\} $
   & 29.73  & $ \left. \begin{array}{cccc} \, \\ \,  \\ 36.61\\ \,\end{array} \right\} $
   & 36.61 & $ \left. \begin{array}{cccc} 2.93 \\ 7.17\\ 8.01\\ 0.89 \end{array} \right\} $ & 19.00 & 5.47\\
   \hline
   $3_{1  3}$    & $ \begin{array}{cc} 2 \\ 3\\ \end{array}$ &  &   &  &  & $ \left. \begin{array}{cc} 0.20 \\ 2.79 \end
      {array} \right\} $ & 2.98 & 2.65\\
   \hline
   $4_{0  4}$    & $ \begin{array}{cc} 0 \\ 3\\ \end{array}$ & $ \left. \begin{array}{cc} 0.01 \\ 3.98 \end{array}
   \right\} $ &  3.99 &  &  & $ \left. \begin{array}{cc} 2.12 \\ 5.56 \end{array} \right\} $ & 7.68 & 5.43\\
   \hline
   $4_{1  3}$    & $ \begin{array}{ccc} 1\\2\\ 3\\ \end{array}$ & $ \left. \begin{array}{ccc} \\2.17\\8.35 \end{array}
   \right\} $  & 10.52  &  $ \left. \begin{array}{ccc} \\ \\13.49 \end{array} \right\} $ & 13.49  & $ \left. \begin
         {array}{ccc} 2.12 \\ 2.60\\2.87 \end{array} \right\} $ & 7.58 & 2.54\\
   \hline
   $5_{0  5}$    & $ \begin{array}{ccc} 1\\2\\ 4\\ \end{array}$ &   &   &   &   & $ \left. \begin{array}{ccc} 0.19 \\
         3.24\\3.30 \end{array} \right\} $ & 6.73 & 4.80\\
   \hline
   $5_{1  4}$    & $ \begin{array}{cc} 0 \\ 4\\ \end{array}$ & $ \left. \begin{array}{cc}  \\ 1.66 \end{array} \right\} $
   &  1.66 &  &  & $ \left. \begin{array}{cc} 0.08 \\ 1.83 \end{array} \right\} $ & 1.90 & 2.22\\
   \hline
   $5_{2  3}$    & $ \begin{array}{cc} 3 \\ 4\\ \end{array}$ & $ \left. \begin{array}{cc} 0.06 \\ 5.61 \end{array}
   \right\} $ &  5.67 &  &  &  &  & 1.92\\
   \hline
   $6_{1  6}$    & $ \begin{array}{cc} 2 \\ 3\\ \end{array}$ &  &  &  &  &  $ \left. \begin{array}{cc} 0.19 \\ 1.07 \end
      {array} \right\} $ & 1.26 & 1.91\\
   \hline\hline
   $\Sigma$&&&99.41&&100.00&&96.67&41.46 \\
\end{tabular}
\caption{Relative population of individual quantum states in the deflected part of the molecular
   beam profile for $T_{\text{rot}}=1.05$~K. Left: IB in He at  1~\% of peak intensity of
   undeflected beam. Center: IB in Ne at  1~\% of peak intensity of undeflected beam. Right: IB in
   Ne at  9~\% of peak intensity of undeflected beam (here orientation images were taken). $P_M$
   denotes relative population of individual $M$-sublevels in~\%, $P_{J_{K_aK_c}}$ the sum over all
   $M$-sublevels, and $P_{J_{K_aK_c}}^{\text{free}}$ the relative population of a given rotational
   quantum state in a free jet.}
\label{tab:results:deflection:pop}
\end{table*}%
For comparison, also the population of each rotational quantum state in the undeflected beam for
this rotational temperature is given. At the position of the laser focus for the orientation
experiments in Ne (column four in \autoref{tab:results:deflection:pop}), the population of the
lowest quantum states is significantly enhanced in the deflected beam compared to the undeflected beam. The fraction of
ground state molecules is enhanced by a factor of five, for instance. About 97~\% of the population
resides in the quantum states listed in \autoref{tab:results:deflection:pop} with the
$J_{K_aK_c}=3_{03}$ state being most abundant. Moving the laser focus even closer towards the upper
cut-off in the beam profile should reduce the number of quantum states that are probed even further.
If a reduction of the beam intensity by two orders of magnitude (compared to the undeflected beam)
can be afforded, only 4 quantum states are predicted to be probed with 37~\% of the molecules being
in the $J_{K_aK_c}=3_{03}$ state. At first glance, it is surprising that this state and not the
absolute ground state, which is expected to have the largest $\mu_{\text{eff}}$, is populated most
in the deflected beam. In order to understand this, the Stark curves for the most deflected quantum
states of IB are shown in \autoref{fig:results:deflection:starkIB}~a), together with their effective
dipole moments (\autoref{fig:results:deflection:starkIB}~b).
Below the relevant electric field strengths, both the $J_{K_aK_c}M = 3_{03}2 $ and the $J_ {K_aK_c}M =
4_{13}3 $ M-sublevels have avoided crossings with close-by states of the same symmetry (dashed lines
in \autoref{fig:results:deflection:starkIB}~a). These avoided crossings lead to large local
effective dipole moments that are comparable to the ground-state $\mu_{\text{eff}}$.~\footnote{On the
   other hand, the local effective dipole moments of the neighboring states (dashed lines in
   \autoref{fig:results:deflection:starkIB}\,a) are reduced by these crossings. At an avoided
   crossing, the two levels that are involved ``exchange'' their effective dipole moments. For large
   asymmetric top molecules, many avoided crossings can lead to a complicated shape of the adiabatic
   Stark curve and a strongly varying effective dipole moment with the electric field strength.}
Thus molecules in these quantum states are deflected as much as ground-state molecules. Furthermore,
states with $M \neq 0$ are doubly degenerate, whereas the ground state with $M = 0$ is only singly
degenerate. Therefore, the population of molecules in the $J_{K_aK_c}M = 3_{03}2 $ in the
undeflected beam is already larger than the population in the ground state.~\footnote{In principle,
   deflection of the quantum states that are coincidentally polar for the range of electric field
   strengths in the deflector could be reduced by operating the deflector at a different high
   voltage. However, due to the large number of quantum states populated and the resulting large
   number of avoided crossings, other quantum states might be conincidentally polar for these
   operation conditions.} From \autoref{tab:results:deflection:pop} it is clear, that it
will be difficult to isolate the rotational ground state of iodobenzene in our setup. Nevertheless,
given that the fraction of ground-state molecules could be increased from 1~\% in the undeflected to 26~\%
in the deflected beam, dramatic effects are to be expected for a variety of further experiments.

We point out, that we are assuming adiabatic following of potential energy curves in all
simulations. Non-adiabatic transitions are unlikely in the strong fields inside the deflector, since
the number of avoided crossings and their energy gaps generally increase with electric field strength.
Moreover, the probability for non-adiabatic following depends on the rate of
change of the field strength, which is only due to the slow translational motion of the molecules.
However, non-adiabatic transitions have been observed in different Stark decelerator beamlines at
real~\cite{Wohlfart:PRA78:033421} and avoided crossings~\cite{Kirste:PRA79:051401} for small
electric fields. Similarly, when the deflected molecules in the experiments reported here enter a
field-free region, scrambling of population over the various $M$ components of their rotational
state will occur.

For small molecules, like OCS or ClCN, the preparation of an ensemble of molecules, all in a single
quantum state will be feasible with the present setup. For these systems, the number of quantum
states that are populated in a supersonic jet is significantly smaller compared to large asymmetric
top molecules like IB or BN. The spacing between neighboring quantum states is larger and the number
of avoided crossings smaller. Thus, the differences in the effective dipole moment between
individual quantum states are larger and, therefore, the degrees of their deflection will vary
considerably.

\subsection{Laser-Induced Alignment of Quantum-State-Selected Molecules}
\label{ssec:results:alignment}

\begin{figure}
   \centering
   \includegraphics[width=\figwidth]{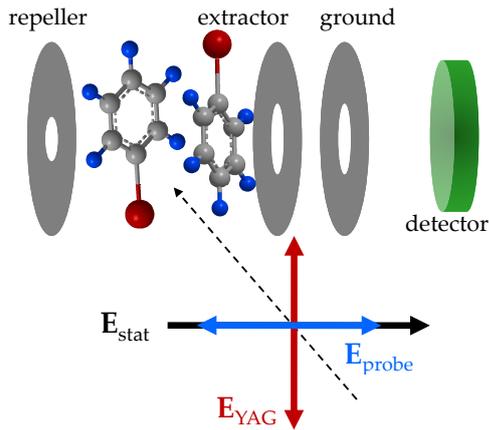}
   \caption{(Color online) Schematic illustration of the polarization state of the YAG and the probe
      pulse with respect to the static electric field and the detector plane used to characterize
      alignment. The dashed line represents the propagation directions of the laser beams. Included
      is also a sketch of the resulting molecular alignment. Repeller, extractor and ground refers
      to the electrostatic plates of the velocity map imaging spectrometer.}
   \label{fig:results:alignment:probegeometry}
   \centering
\end{figure}

\begin{figure}
   \includegraphics[width=\figwidth]{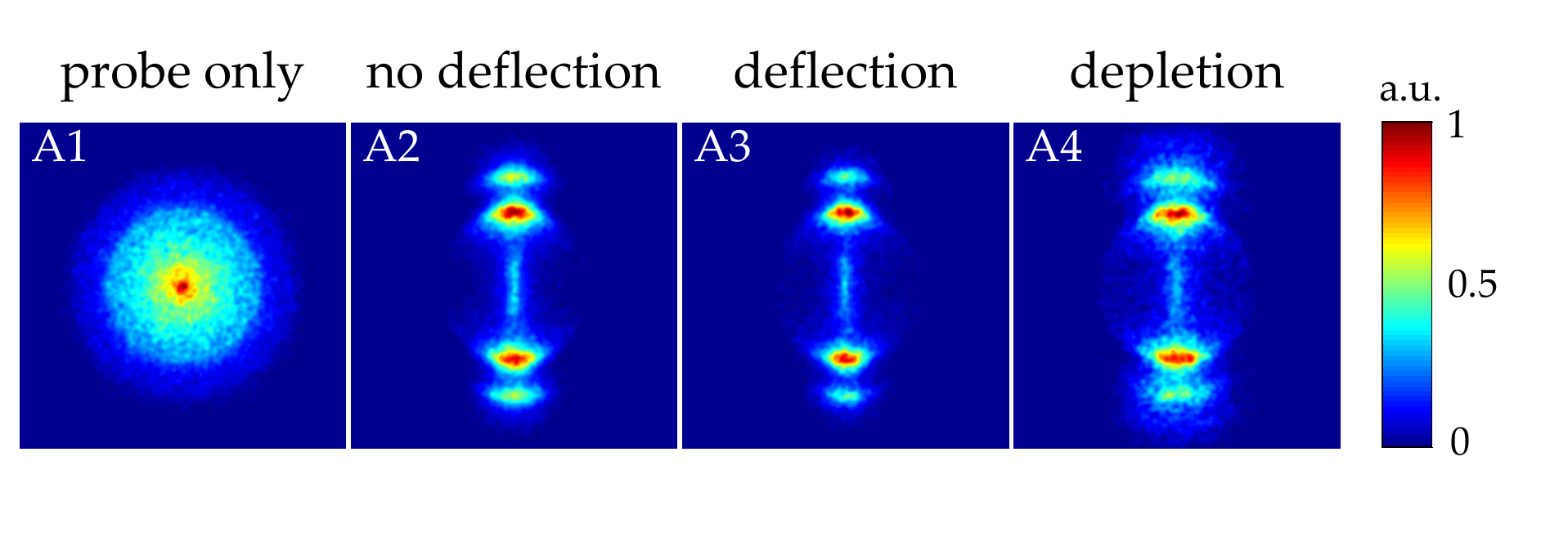}
   \caption{(Color online) I$^+$ ion images illustrating alignment, recorded when the probe pulse
      Coulomb explodes the iodobenzene molecules seeded in He. The polarizations of the YAG and the
      probe pulses are kept fixed as illustrated in \autoref{fig:results:alignment:probegeometry}.
      The labels ``no deflection'', ``deflection'' and ``depletion'' correspond to images recorded
      at lens position $y=0.0$~mm, 1.0~mm and -0.9~mm respectively, the latter two with the deflector
      at 10~kV (see \autoref{fig:results:deflection:IB}). The intensities of the YAG and probe pulse
      are $8\times 10^{11}$~W/cm$^2$ and $5\times10^{14}$~W/cm$^2$, respectively. The color scale indicates the relative number of ions. This color scale is the same for all subsequent figures
      showing ion images.}
   \label{fig:results:alignment:images:helium}
\end{figure}

We now turn to studying alignment induced by the YAG pulse. The basic experimental observables are
2D I$^+$ ion images recorded when the iodobenzene molecules are irradiated with both the YAG pulse
and the probe pulse. The geometry of the laser pulse polarizations with respect to the velocity map
imaging spectrometer (VMI) is illustrated in \autoref{fig:results:alignment:probegeometry}.
The YAG pulse is linearly polarized along the vertical direction, \ie, in the detector plane. The
probe pulse is linearly polarized perpendicular to the detector plane, which ensures that there is
no detection bias on the molecular orientation in that plane. This results in a circularly symmetric
I$^+$ image, when only the probe pulse is used (\autoref{fig:results:alignment:images:helium} A1).
When the YAG pulse is included (\autoref{fig:results:alignment:images:helium} A2-A4) the I$^+$
images exhibit strong angular confinement along the polarization of the YAG pulse.

The I$^+$ ions appear as two pairs of radially localized regions,
corresponding to two different fragmentation channels of the Coulomb
explosion. The radius of the outermost (and weakest) pair of rings
is approximately $\sqrt{2}$ times larger than the radius of the
innermost (and brightest) pair of rings. Since the radius is
proportional to the velocity of the ions the I$^+$ ions from the
outermost pair of rings originate from a Coulomb explosion channel
that releases twice as much kinetic energy as the channel producing
the I$^+$ ions in the innermost pair of rings. As pointed out in
several previous studies from our group (see for instance
reference~\,\onlinecite{Larsen:JCP111:7774}) this is only consistent with the
innermost pair of rings originating from iodobenzene being doubly
ionized by the probe pulse and fragmenting into an I$^+$+
C$_6$H$_5$$^+$ ion pair, and the outermost pair of rings originating
from I$^+$ ions formed by triple ionization and fragmentation into
an I$^+$+ C$_6$H$_5$$^{2+}$ ion pair. The pronounced angular
confinement observed in images A2-A4 is quantified by calculating
the expectation value of \cost, where $\theta_{2D}$ is the angle
between the YAG pulse polarization and the projection of the I$^+$
recoil velocity vector onto the detector plane. In this paper,
\cost\ values are calculated only from ions detected in radial
region corresponding to the I$^+$+ C$_6$H$_5$$^{2+}$ channel. By
doing so, the YAG intensities probed are restricted to a narrow
range close to the maximum value, as the high nonlinearity of the
multiphoton process occurs efficiently only in the spatial regions
close to the focal point of the YAG beam.

Image A1, recorded with only the probe pulse present, should correspond to a target of randomly
oriented molecules. As expected the image is circular symmetric and $\cost=0.515$. When the YAG
pulse is included (image A2) a pronounced angular confinement is observed along the polarization of
the YAG pulse and \cost is increased to 0.947. These observations are in complete agreement with
previous studies~\cite{Kumarappan:JCP125:194309}. When the deflector is turned on and the laser foci
moved to the edge of the most deflected molecules (at the position marked in
\autoref{fig:results:deflection:IB} a), corresponding to molecules in the lowest rotational states,
the angular confinement is further enhanced (image A3) leading to a \cost\ value of 0.968. By
contrast, when the experiment is conducted on the least deflected molecules in the depleted region
(image A4), corresponding to molecules in the highest rotational states, the alignment is weakened
and \cost=0.900.

\begin{figure}
   \centering
   \includegraphics[width=0.8\figwidth]{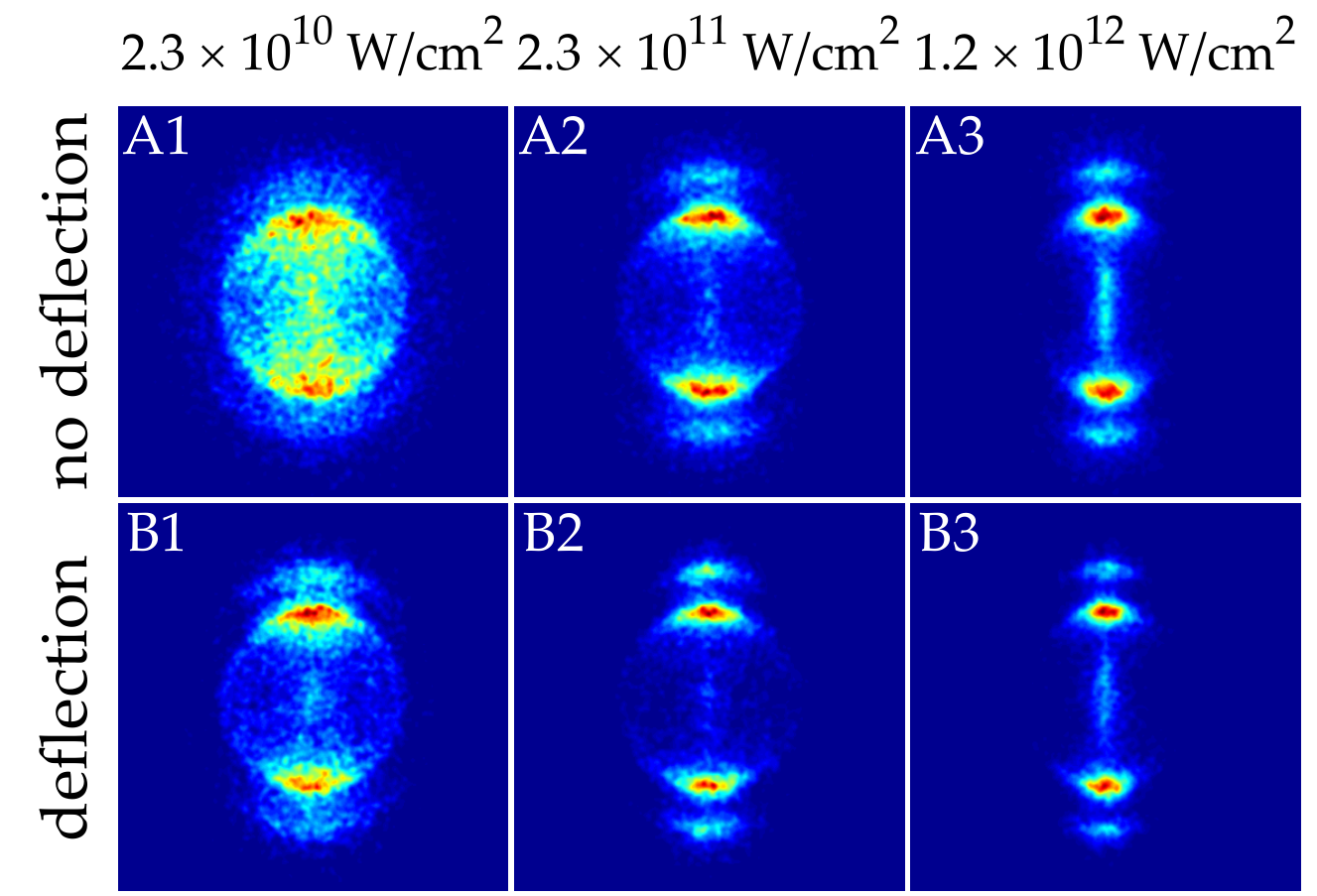}
   \caption{(Color online) I$^+$ion images illustrating alignment at different intensities of the
      YAG pulse, recorded when the probe pulse Coulomb explodes iodobenzene molecules seeded in 20
      bar Ne. The labels "no deflection" and "deflection" correspond to images recorded at lens
      position $y=0.0$~mm (deflector turned off) and 2.15~mm (deflector at 10~kV), respectively. The
      intensity of the probe pulse is $5\times10^{14}$~W/cm$^2$.}
   \label{fig:results:alignment:images:neon}
\end{figure}

We repeated the alignment measurements when iodobenzene was seeded in Ne. The results are displayed
in \autoref{fig:results:alignment:images:neon}.
Like in the He case a pronounced improvement is observed when deflected rater than undeflected
molecules are employed. \autoref{fig:results:alignment:images:neon} shows images of I$^+$ recorded
at three different intensities of the YAG laser for both undeflected and deflected molecules seeded
in Ne. The effect of the deflector is clearly seen when comparing, for instance, image B1 (deflected)
and A1 (undeflected). At this low YAG intensity ($2.3\times10^{10}$~W/cm$^2$) weak alignment is
obtained in the non-deflected beam with \cost$=0.695$. Going to the edge of the deflected molecular
beam (position indicated in \autoref{fig:results:deflection:IB} b ) a clear enhancement is observed
(image B1) with the \cost\ value rising to 0.869. Also, at high YAG intensity
($1.2\times10^{12}$~W/cm$^2$) the difference in angular confinement comparing the undeflected (image
A3) to the deflected molecules (image B3) is visible. While \cost=0.929 represents the limit
of the degree of alignment of iodobenzene seeded in Ne in the undeflected beam, emloying the
deflector leads to an unprecedented degree of laser-induced alignment of $\cost=0.972$. We note that
$\langle\cos^2\theta_{2D}\rangle$~=~1 would correspond to the quantum mechanically unfeasible
situation of perfectly 1D aligned molecules.

\begin{figure}
   \centering
   \includegraphics[width=\figwidth]{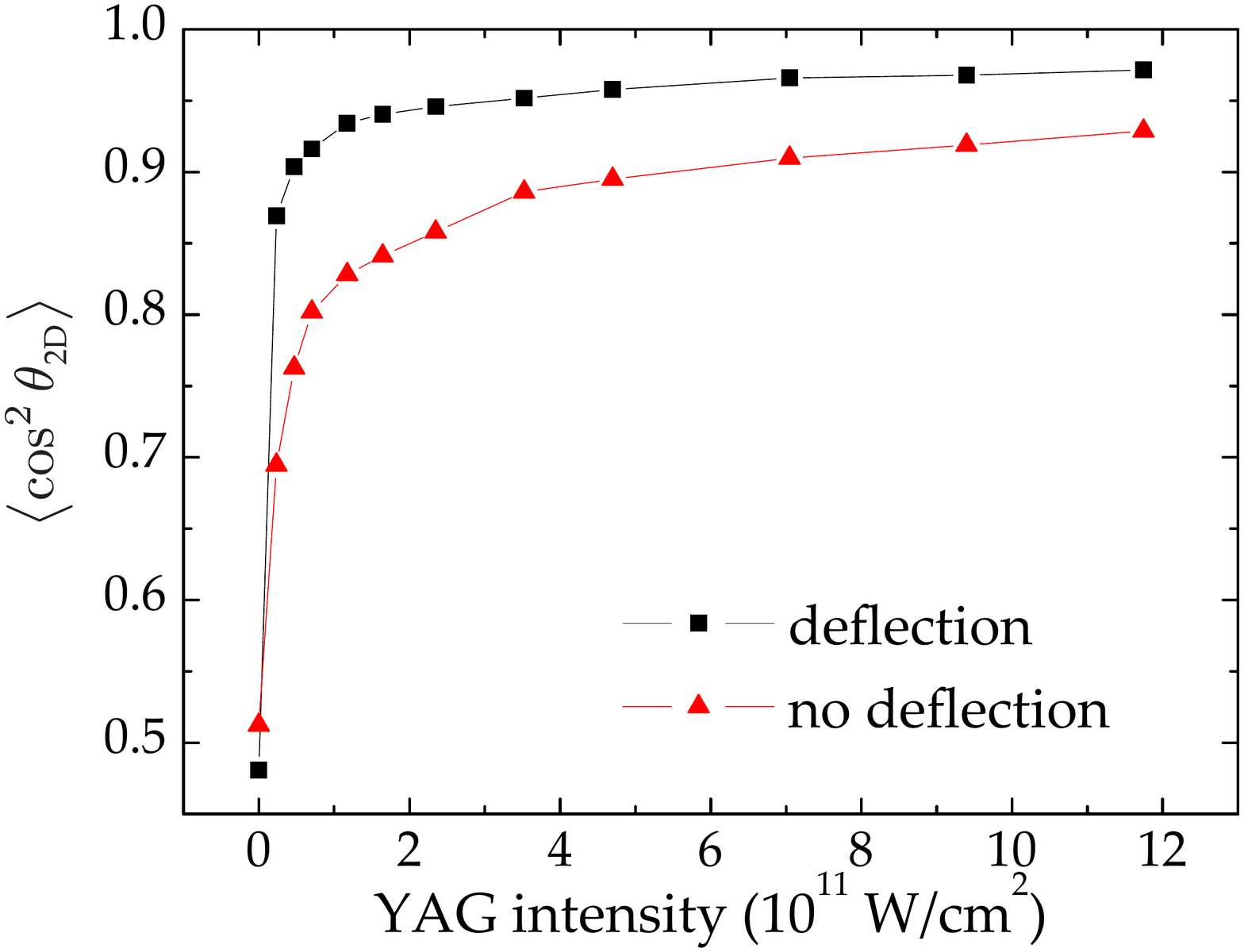}
   \caption{Degree of alignment as a function of the YAG intensity for iodobenzene seeded in 20 bar
      Ne. The labels "no deflection" and "deflection" correspond to images recorded at lens
      position $y=0.0$~mm (deflector turned off) and 2.15~mm (deflector at 10kV), respectively. The
      intensity of the probe pulse is $5\times10^{14}$~W/cm$^2$.}
   \label{fig:results:alignment:intensity}
\end{figure}

To quantify the angular information of the images, \cost\ values are plotted as a function of YAG
intensity, the results are displayed in \autoref{fig:results:alignment:intensity}.
Even at very low laser intensities a high degree of alignment can be obtained from an ensemble of
quantum-state-selected molecules. The tendency shown in this graph with a steep rise and early
saturation of the degree of alignment agrees with previous results investigating the dependence of
alignment on the rotational temperature of the ensemble of
molecules~\cite{Kumarappan:JCP125:194309}. Effectively, the quantum-state selection corresponds to a ``colder'' albeit non-thermal beam (see \autoref{ssec:results:deflection}).

Note that the contrast between the undeflected and the deflected beam is expected to be greater if
Ne is used instead of He as a carrier gas. In the undeflected beam the maximum degree of alignment
that can be achieved is smaller in Ne because the rotational cooling in the supersonic expansion is
less effective due to the lower stagnation pressure\cite{Hillenkamp:JCP118:8699}. Additionally, in
the deflected beam a better degree of alignment is expected for Ne, as the efficiency of the
quantum-state selection in the present setup is significantly enhanced due to the longer residence
time in the deflector (see \autoref{ssec:results:deflection}).


\subsection{Laser-Induced Orientation of Quantum-State-Selected Molecules}
\label{ssec:results:orientation}

\begin{figure}
   \centering
   \includegraphics[width=\figwidth]{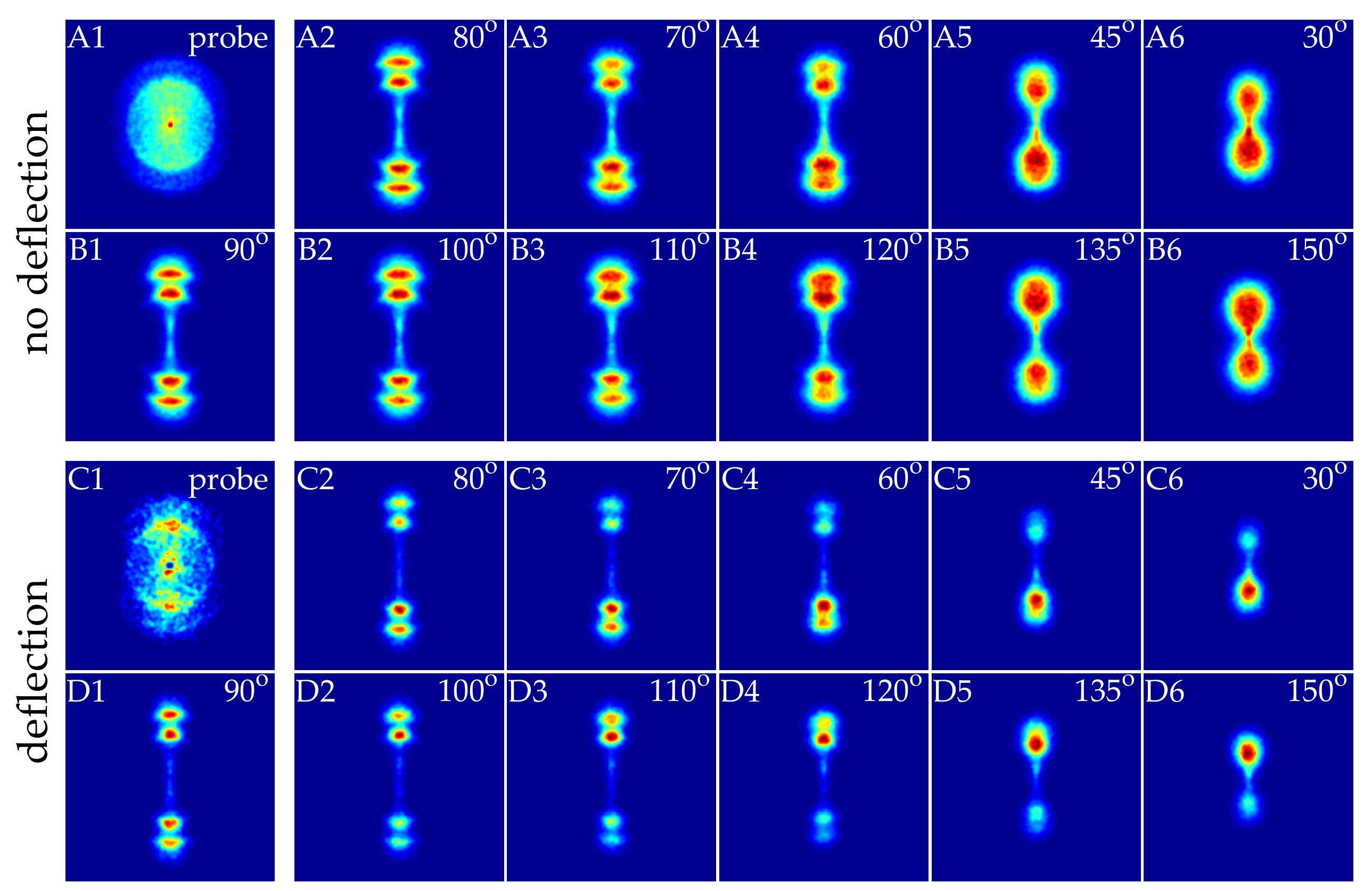}
   \caption{(Color online) I$^+$ ion images illustrating orientation for different values of
      $\beta$, recorded when the circularly polarized probe pulse Coulomb explodes the iodobenzene
      molecules seeded in 90 bar He. The labels ``no deflection'' and ``deflection'' correspond to
      images recorded at lens position $y=0.0$~mm (deflector turned off) and 1.0~mm (deflector at
      10~kV), respectively. The intensity of the YAG and the probe pulse is $8\times
      10^{11}$~W/cm$^2$ and $5\times10^{14}$~W/cm$^2$, respectively. E$_{\text{stat}}$~$=594$~V/cm.}
   \label{fig:results:orientation:series:helium}
\end{figure}

Next, we discuss orientation due to the combined action on the molecules by the YAG pulse and the
static electric field (E$_{\text{stat}}$) from the VMI electrodes~\cite{Friedrich:JCP111:6157,
   Friedrich:JPCA103:10280}.
\begin{figure}
   \centering
   \includegraphics[width=\figwidth]{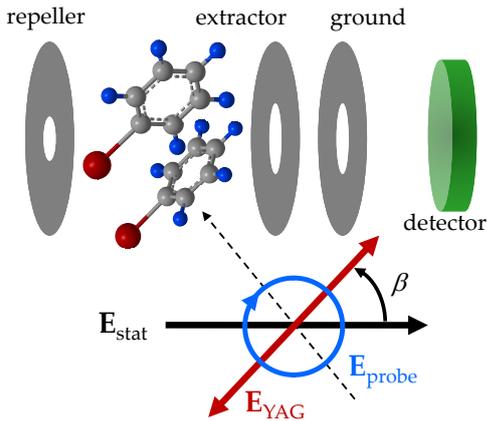}
   \caption{(Color online) Schematic illustration of the polarization state of the YAG and the probe
      pulse with respect to the static electric field and the detector plane used to characterize
      orientation. The dashed line represents the propagation direction of the laser beams. Also, a
      sketch illustrating the molecular orientation is included.}
   \label{fig:results:orientation:probegeometry}
\end{figure}
\autoref{fig:results:orientation:probegeometry} illustrates the polarization state of the YAG and
the probe pulse with respect to the static electric field of the VMI electrodes. The important
difference compared to the alignment data is that the YAG polarization is
rotated away from the axis perpendicular to the static field. Thus, the orientation data result
from geometries where the angle $\beta$, between the YAG polarization (the C-I bond axis) and the
static electric field, is different from 90\degree - see
\autoref{fig:results:orientation:probegeometry}. To image orientation a circularly polarized probe
pulse is used. This ensures that any molecule will be ionized - and thus detected - with the same
probability independent of $\beta$. This circularly polarized probe will induce some bias on the
angular distribution of the I$^+$ ions (see \autoref{fig:results:orientation:series:helium} A1 and
C1), however, importantly, it is up/down symmetric.
\autoref{fig:results:orientation:series:helium} shows I$^+$ ion images for different $\beta$ values
for both deflected and undeflected molecules seeded in He. As mentioned the circularly polarized
probe alone gives rise to an image that exhibits some angular confinement with \cost=0.70
(\autoref{fig:results:orientation:series:helium} A1). Consequently, including the YAG pulse at
$\beta=90$\degree\ results in an image (\autoref{fig:results:orientation:series:helium} B1) that
appears slightly different from the corresponding image with a linearly polarized probe
(\autoref{fig:results:alignment:images:helium} A2), but still shows that the molecules are tightly
aligned.

Focussing first on the non-deflected data of \autoref{fig:results:orientation:series:helium} (row A
and B) two prominent changes are observed as the polarization of the YAG pulse is gradually rotated
away from the detector plane (images A2-A6 and B2-B6). First, the location of the I$^+$ rings shifts
closer to the center of the images. This is due to the fact that the C-I axis alignment, and thus
the emission direction of the I$^+$ ions, follows the YAG pulse polarization. When the C-I axis is
aligned at an angle $\beta$ the magnitude of the I$^+$ velocity vector recorded on the detector will
be reduced by the factor $\sin(\beta)$. The detrimental effect on the radial (velocity) resolution
is obvious at $\beta=135\degree/45\degree$ (image A5 and B5) and $30\degree/150\degree$ (image A6
and B6), where the two I$^+$ explosion channels, I$^+$+ C$_6$H$_5$$^{+}$ and I$^+$+
C$_6$H$_5$$^{2+}$, become indistinguishable as they merge in the 2D projection
onto the detector plane.

Secondly, as the YAG pulse polarization is turned away from 90\degree the up/down symmetry of the
images, characteristic for the alignment data described in \autoref{ssec:results:alignment} (and
\autoref{fig:results:orientation:series:helium} column 1), is broken. For images with
$90\degree<\beta<180\degree$ (images B2-B6) more I$^+$ ions are detected in the upper part, whereas
for $0\degree{}<\beta<90\degree$ (images A2-A6) more I$^+$ ions are detected in the lower part. The
asymmetry becomes more pronounced as the YAG polarization is rotated closer to the axis of the
static field. We interpret these observations as orientation due to the combined effect of the YAG
laser field and the projection of the static electric extraction field (E$_{\text{stat}}$) on the
YAG polarization axis. This projection (numerical value: $\mid\cos(\beta)\mid\cdot\,\text{E}_{\text{stat}}$
increases as $\beta$ is rotated towards 0\degree\ or 180\degree, which is expected to cause an
increase of the orientation~\cite{Friedrich:JCP111:6157, Friedrich:JPCA103:10280}, in agreement with
the experimental findings.

As discussed in \autoref{ssec:results:deflection} all states of iodobenzene are high-field seeking,
hence the orientation is expected to place the I-end of the molecules towards the repeller plate
(see \autoref{fig:results:orientation:probegeometry}), where the electrical potential is highest,
because the dipole moment of iodobenzene is directed along the C-I axis pointing from iodine
(``negative end'') towards the phenyl ring (``positive end''). The expected resulting molecular
orientation at a given angle of $\beta$ is shown in \autoref{fig:results:orientation:probegeometry}.
Thus, for $0\degree<\beta<90\degree$ the I$^+$ ions are expected to preferentially be ejected
downwards, and for $90\degree<\beta<180\degree$ they will be ejected upwards. This is in agreement
with the up/down asymmetry in the images.

The alignment (images D1) and orientation (images C2--C6 and D2--D6) improves significantly when the
deflector is turned on and the foci of the lasers are moved to the position of the most deflected
molecules [position marked in \autoref{fig:results:deflection:IB}~a)]. The markedly better
orientation resulting in a much more pronounced up/down asymmetry is clearly visible even when the
the YAG pulse is only turned slightly away from perpendicular, \ie, comparing deflected and
undeflected images for $\beta$=100\degree\ (image C2 and A2) and $\beta=80\degree$ (image D2 and
B2). From the previous discussion it appears that the highest degree of orientation is achieved when
$\beta$ is rotated towards 0\degree\ or 180\degree. This is clearly seen from the images in row C
and D and, again, the improvement obtained with deflected molecules is striking - compare image C6
to A6 (or D6 to B6).

\begin{figure}
   \centering
   \includegraphics[width=\figwidth]{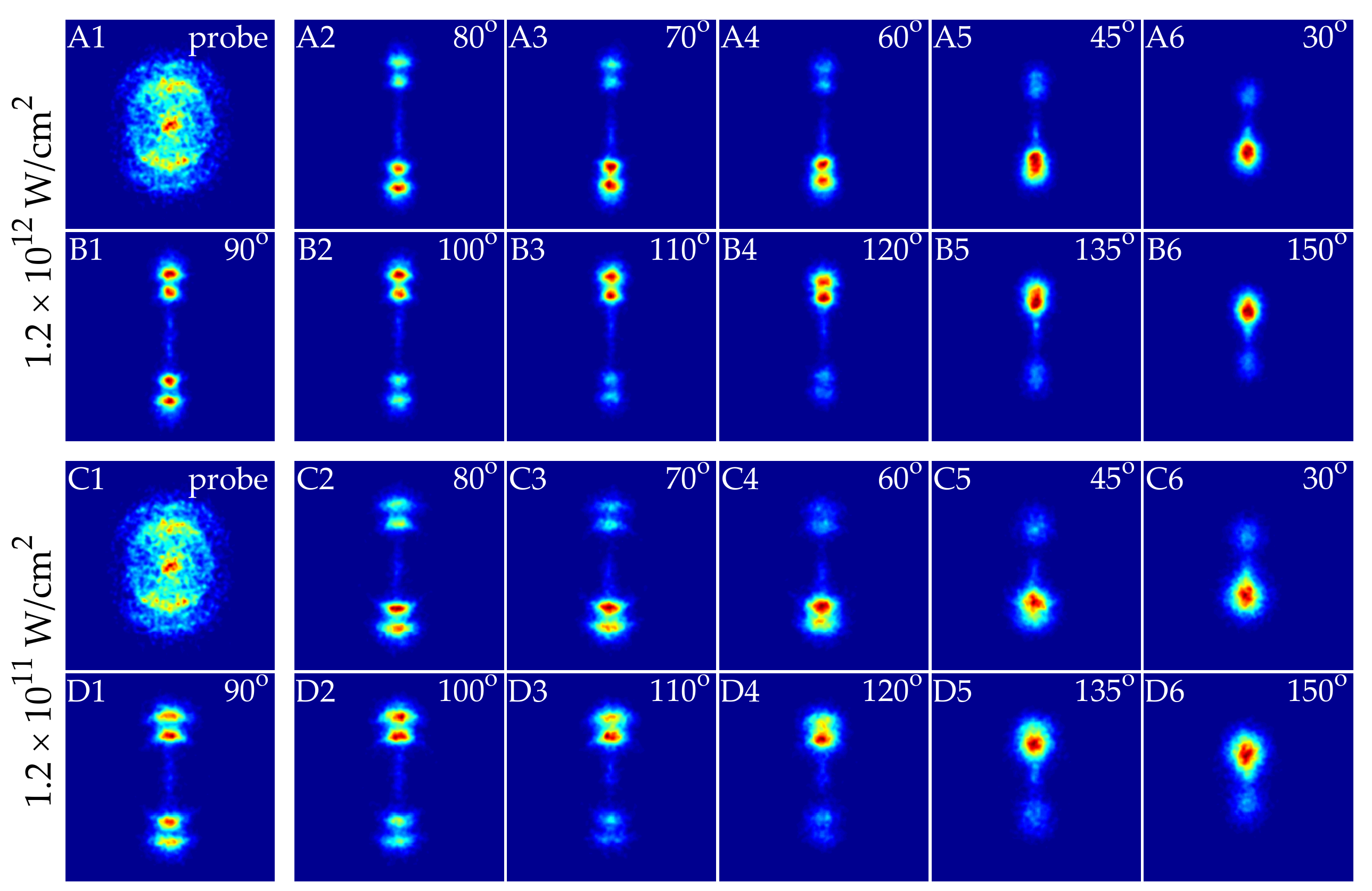}
   \caption{(Color online) I$^+$ion images of deflected iodobenzene seeded in 20 bar Ne, illustrating
      orientation at two different intensities of the YAG pulse. The images are recorded at lens position
       $y=2.25$~mm with the deflector at 10~kV. The intensity of the probe pulse is $5\times10^{14}$~W/cm$^2$.
       E$_{\text{stat}}$~$=594$~V/cm.}
   \label{fig:results:orientation:series:neon}
\end{figure}

Similar orientation measurements were conducted for iodobenzene seeded in Ne instead of in He.
\autoref{fig:results:orientation:series:neon} shows I$^+$ images at a series of $\beta$ values for
two different intensities of the YAG pulse recorded with the deflector at 10~kV at the position
marked in \autoref{fig:results:deflection:IB}\,b).
Compared to the images displayed in \autoref{fig:results:orientation:series:helium} row A and B,
although recorded at slightly different intensities of the YAG pulse, a significant improvement is
observed. Furthermore even at low intensity of the YAG a high degree of orientation for iodobenzene
seeded in Ne is achieved. At the same time some loss in the angular confinement, \ie, in the
alignment degree, is visible. We assign the clear improvement in the up/down asymmetry to the more
stringent state selection in Ne compared to He as described in \autoref{ssec:results:deflection},
hence, manifesting itself in a higher degree of orientation.

\begin{figure}
   \centering
   \includegraphics[width=\figwidth]{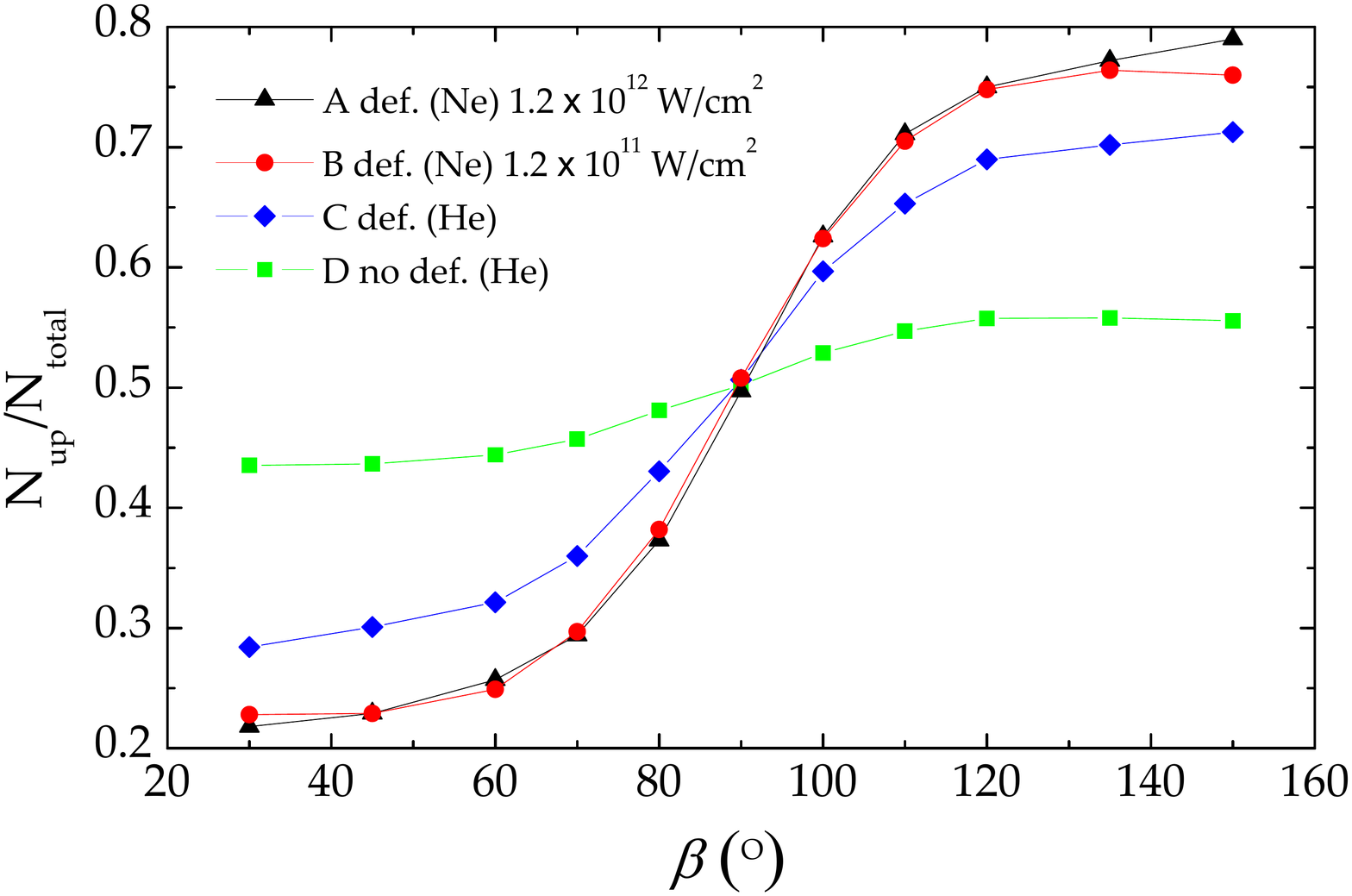}
   \caption{Orientation of iodobenzene, seeded in either He or Ne, represented by the number of I$^+$ ions in the upper half of
      the image (N$_{\text{up}}$) divided by the total number of I$^+$ ions in the image
      (N$_{\text{total}}$) as a function of $\beta$. For the experiments
      conducted with He as a carrier gas, curves C (lens position $y=1.0$~mm and deflector at 10~kV)
      and D (lens position $y=0.0$~mm and deflector off), the intensity of the YAG pulse was
      $7.8\times10^{11}$~W/cm$^2$. For the Ne carrier gas, curves A and B (lens position $y=2.25$~mm
      and deflector at 10~kV), the intensities of the YAG pulse are displayed in the insert. The
      intensity of the probe pulse for all curves was fixed at $5\times10^{14}$~W/cm$^2$,
      E$_{\text{stat}}=594$~V/cm.}
   \label{fig:results:orientation:up/down}
\end{figure}

To quantify the up/down asymmetry, \ie, the degree of orientation, we determine for each image the
number of I$^+$ ions, N(I$^+$)$_\text{up}$, in the upper part of the I$^+$~+~C$_6$H$_5$$^{+}$ and
I$^+$~+~C$_6$H$_5$$^{2+}$ channels (\ie, ions detected in the upper half of the images) as well as
the total number of ions,
$\text{N}(\text{I}^+)_\text{total}\left(=\text{N}(\text{I}^+)_\text{up}+\text{N}(\text{I}^+)_\text{down}\right)$.
This ratio, as a function of $\beta$, is displayed in \autoref{fig:results:orientation:up/down}.
Focussing first on curves C and D, representing iodobenzene seeded in He, the difference between the
data for the deflected molecules and the data obtained with the deflector turned off is striking and
shows the advantage of selecting the lowest-lying rotational states for strongly increasing the
degree of orientation. The further improvement when Ne is used instead of He is clear from curves A
and B. These two curves also show that the pronounced degree of orientation is maintained when the
intensity of the YAG pulse is lowered by an order of magnitude compared to the maximum value of
$1.2\times10^{12}$~W/cm$^2$.

\section{Conclusion and Outlook}
\label{sec:conclusion-outlook}

In conclusion, we have shown that deflection of cold molecular beams with an inhomogeneous static
electric field enables the selection and the spatial separation of the most polar quantum states,
\ie, the lowest-lying rotational states. The method demonstrated here is complementary to
state-selection for small molecules using a hexapole focuser, which has been suggested to be applied
for improved alignment and orientation experiments~\cite{Gijsbertsen:PRL99:213003} and recently been
experimentally demonstrated~\cite{Ghafur:NatPhys5:289}. While a hexapole focuser only works for small
molecules in low-field-seeking quantum states, beam deflection will apply broadly to a wide range of
molecules, from diatomics to large biomolecules. The deflection is strongest for molecules with a
large permanent dipole moment to mass ratio. For a given molecule the deflection is optimized by
employing stronger deflection fields, increasing the length of the deflector, or lowering the speed
of the molecule, for instance, by using neon rather than helium as a carrier gas. For small
molecules, the preparation of an ensemble of molecules all in a single quantum state should be
feasible. As an application of the state-selected molecules we showed that selection of iodobenzene
in low lying rotational states allows to achieve unprecedented degrees of laser-induced adiabatic
alignment and mixed laser- and static-field orientation. In particular, we demonstrated that strong
alignment and orientation can be maintained even when the intensity of the alignment pulse is
lowered to the $10^{10}$--$10^{11}$~W/cm$^2$ range. This can reduce unwanted disturbance from the
laser field in future applications of adiabatically aligned or oriented molecules. We note that it
should be possible to improve the degree of orientation obtained here simply by increasing the
static electric field. Due to experimental constraints this was not implemented in the present work.\\

Getting access to cold molecules in the gas phase typically involves using a molecular beam from a
supersonic expansion that usually consists of more than 99 percent carrier gas and less than one
percent of the specific molecules. In several types of experiments the atomic carrier gas can
contribute to, or even completely overshadow the particular signal measured. The electrostatic
deflection naturally separates the polar molecules from the unpolar carrier gas and thus removes this
unwanted background. We conservatively estimate  the density in the original molecular beam to be $10^{11}$ molecules/cm$^{3}$. In the experiments presented here the density in the deflected part of the beam is approximately $10^{10}$ molecules/cm$^3$. We foresee this density to be sufficient for a variety of applications, such as photoelectron spectroscopy with VUV, EUV~\cite{Ng:ARPC53:101}, or x-ray light sources, including
attosecond pulses, or high harmonic generation experiments with fs laser
pulses~\cite{Li:Science322:1207}. For all of these applications the separation of the molecular
target from the carrier gas might be of great relevance.\\

Additionally,  brute-force orientation~\cite{Loesch:JCP93:4779, Friedrich:Nature353:412}, that is, the spatial
orientation of polar molecules using strong dc electric fields, will benefit from the state-selected
samples similar to what was demonstrated here. In fact, the states that are deflected the most are
also oriented the most in a dc electric field. To illustrate the achievable
orientation, we have calculated the ensemble averaged orientation in a homogeneous electric field of
250~kV/cm for iodobenzene for a thermal ensemble of 1~K. For this ensemble
$\left<\cos\theta\right>=0.757$ is obtained. For a deflected, quantum-state-selected sample of
iodobenzene molecules at 1~\% of the undeflected peak intensity (see
\autoref{tab:results:deflection:pop}), an increased ensemble averaged orientation of
$\left<\cos\theta\right>=0.905$ is predicted.
For nonadiabatic alignment~\cite{RoscaPruna:PRL87:153902,Stapelfeldt:RMP75:543} it was demonstrated, that the dynamics and, importantly, the degree of
alignment and orientation depends strongly on the initial rotational state
distribution~\cite{Seideman:JCP115:5965, Machholm:JCP116:10724, RoscaPruna:JCP116:6579,
   Peronne:PRA70:063410, Muramatsu:PRA79:011403}. Selection of rotational states is, therefore,
highly advantageous for nonadiabatic laser-induced schemes to control the spatial orientation of
molecules, as recently demonstrated for NO molecules.~\cite{Ghafur:NatPhys5:289}
In general, complete elimination of the rotational tumbling of an asymmetric top molecule requires,
that all three principal axes are confined along laboratory fixed axes. This is the area of 3D
alignment and orientation, subjects that have been treated theoretically and
experimentally~\cite{Larsen:PRL85:2470, Tanji:PRA72:063401, Lee:PRL97:173001, Viftrup:PRL99:143602,
   Rouzee:PRA77:043412, Nevo:PCCP:inprep}. A straightforward extension of the current work
would be significantly improved 3D alignment and orientation of polar molecules by employing
state-selected molecules and we have recently performed such
experiments~\cite{Nevo:PCCP:inprep}.

For large (bio-)molecules, typically multiple structural isomers
(conformers)~\cite{Suenram:JACS102:7180} are present even at the low temperatures in a supersonic
jet~\cite{Rizzo:JCP83:4819}. These conformers often exhibit large, and
largely different dipole moments what can be exploited to spatially separate them using inhomogeneous electric fields. This separation has recently been demonstrated in an alternating gradient focusing
(AG) selector for the cis- and trans-conformers of 3-aminophenol~\cite{Filsinger:PRL100:133003}.
However, also the static field of a deflector can be used to spatially isolate individual
conformers~\cite{Filsinger:ACIE48:6900}. In general, the ability to achieve very high degrees of alignment and orientation is of great
interest for a number of applications. We therefore believe state-selected molecules could be very
beneficial for areas such as photoelectron angular distributions from fixed-in-space
molecules~\cite{Bisgaard:Science323:1464}, (ultrafast) diffraction with electron or x-ray
sources~\cite{Spence:PRL92:198102, Peterson:APL92:094106} and time-resolved studies of light-induced stereochemistry~\cite{Madsen:PRL102:073007}. Furthermore,  Janssen and coworkers have shown that the ability to select a single rotational state with a hexapole focuser enables new possibilities for studying directional dynamics of fragments in photodissociation of small molecules~\cite{Rakitzis:Science303:1852}. The deflection method strongly increases the number of molecules to which single rotational state selection, and subsequent orientation, can be applied. In particular, it will offer access to studies of photoinitiated processes in oriented targets of larger asymmetric tops.

\appendix
\section{Simulation of Beam Profiles}

In this Appendix, the simulations of the experimental deflection profiles are described in detail.
First, the calculation of the Stark effect for asymmetric top molecules is explained in
\autoref{appendix:stark}. \autoref{appendix:simulations} describes how single-quantum-state
deflection profiles are obtained from trajectory simulations using the calculated Stark curves.
Finally, the single-quantum-state profiles are averaged in a suitable way to simulate a thermal
ensemble for a given rotational temperature, which is described in \autoref{appendix:profiles}, and
fitted to the experimental data (\autoref{appendix:fit}).

\subsection{Stark Effect Calculations}
\label{appendix:stark}

In order to calculate the adiabatic energy curves for asymmetric top molecules, the Hamiltonian
matrix is set up in the basis of symmetric top wavefunctions. In the presence of an electric field,
only $M$ is a good quantum number for an asymmetric rotor. $J$, which is a good quantum number in
the field-free case, is mixed by the field, whereas $K$ is mixed by the molecular asymmetry. Thus,
the adiabatic energy curves can be calculated for the different $M$ levels individually by setting
up and diagonalizing the $M$ matrices including all $J$ and $K$ levels. An accurate description of
higher rotational quantum states also requires including centrifugal distortion constants. The
Hamiltonian $H$ of an asymmetric rotor molecule with dipole moment $\vec{\mu}$ in an electric field
of strength $E$ can be written as the sum of the Hamiltonian $H_{\text{rot}}$ of an asymmetric rotor
in free space in Watson's A-reduction~\cite{Watson:VibSpecStruct6:1} and the contribution due to the
Stark effect $H_{\text{Stark}}$ as
\begin{equation}
   H = H_{\text{rot}} + H_{\text{Stark}}
\end{equation}
Following references\;\onlinecite{Watson:VibSpecStruct6:1} and\;\onlinecite{AbdElRahim:JPCA109:8507},
the corresponding matrix elements are: \footnote{Note that in
   reference\;\onlinecite{AbdElRahim:JPCA109:8507} representation $I^l$ is used.}
\begin{widetext}
   \vspace{-5mm}
   \enlargethispage{6mm}
   \begin{align}
      \left<JKM \mid H_\text{rot} \mid JKM\right> &= { \frac{B+C}{2} \left(J(J+1)-K^2\right) + AK^2 } \notag\\
      & \quad - \Delta_J J^2(J+1)^2 - \Delta_{JK} J(J+1)K^2 - \Delta_K K^4 \\
      \left<JK\pm2M \mid H_\text{rot} \mid JKM\right> &= { \left(\frac{B-C}{4} -
            \delta_J J(J+1) - \frac{\delta_K}{2}\left((K\pm2)^2 + K^2\right) \right)} \notag\\
      & \quad \cdot \sqrt{J(J+1) - K(K \pm 1)} \sqrt{J(J+1)-(K \pm 1)(K \pm 2)} \\
      \left<JKM \mid \mu_a \mid JKM\right> &= -\frac{MK}{J(J+1)} \mu_a E \\
      \left<J+1KM \mid \mu_a \mid JKM\right> &= \left<JKM \mid \mu_a \mid J+1KM\right> \notag\\
      &= {- \frac{\sqrt{(J+1)^2-K^2} \sqrt{(J+1)^2 - M^2}}{(J+1) \sqrt{(2J+1)(2J+3)}} \mu_a E } \\
      \left<JK\pm1M \mid \mu_b \mid JKM\right> &= {-
         \frac{M \sqrt{(J \mp K)(J \pm K+1)}}{2J(J+1)} \mu_b E }\\
      \left<J+1K\pm1M \mid \mu_b \mid JKM\right> &= \left<JK\pm1M \mid \mu_b \mid J+1KM\right>  \notag\\
      &= {\pm \frac{\sqrt{(J \pm K+1)(J \pm K+2)}\sqrt{(J+1)^2 - M^2}}{2(J+1) \sqrt{(2J+1)(2J+3)} } \mu_b E } \\
      \left<JK\pm1M \mid \mu_c \mid JKM\right> &= {\pm i
         \frac{M \sqrt{(J \mp K)(J \pm K+1)}}{2J(J+1)} \mu_c E }\\
      \left<J+1K\pm1M \mid \mu_c \mid JKM\right> &= \left<JK\pm1M \mid \mu_c \mid J+1KM\right> \notag\\
      &= {-i \frac{\sqrt{(J \pm K+1)(J \pm K+2)} \sqrt{(J+1)^2 - M^2 }}{2(J+1) \sqrt{(2J+1)(2J+3)}} \mu_c E }
   \end{align}
\end{widetext}
For the correct assignment of the states to the ``adiabatic quantum number labels''
$\tilde{J}_{\tilde{K_a}\tilde{K_c}}\tilde{M}$, \ie, to the adiabatically corresponding field-free
rotor states, one has to classify the states according to their character in the \emph{electric
   field symmetry group}~\cite{Watson:CJP53:2210, Bunker:MolecularSymmetry}. This symmetry
classification can be performed by applying a Wang transformation~\cite{Wang:PR34:243} to the
Hamiltonian matrix. If the molecule's dipole moment is along one of the principal axes of inertia,
the matrix will be block diagonalized by this transformation according to the remaining symmetry in
the field and the blocks can be treated independently. For arbitrary orientation of the dipole
moment in the inertial frame of the molecule the full matrix must be diagonalized. In any case, this
process ensures that all states (eigenvalues and eigenvectors) obtained from a single matrix
diagonalization do have the same symmetry, and, therefore, no real crossings between these states
can occur. Therefore, by sorting the resulting levels by energy and assigning quantum number labels
in the same order as for the field-free states of the same symmetry yields the correct adiabatic
labels. These calculations are performed for a number of electric field strengths -- typically in
steps of 1~kV/cm from 0~kV/cm to 200~kV/cm -- and the resulting energies, $W_{\text{Stark}}(E)$, are
stored for later use in simulations using the libcoldmol program package~\cite{Kuepper:libcoldmol}.

\subsection{Monte Carlo Simulations of a Molecular Beam}
\label{appendix:simulations}

Simulations of the electrostatic beam deflection in our setup are performed with the home-built
software package libcoldmol~\cite{Kuepper:libcoldmol}. Trajectories for individual molecules in a
given rotational quantum states are obtained from numerical integration of the 3D equations of
motion using a Runge-Kutta algorithm. The initial phase space distribution of the molecular packet
in the transverse spatial coordinates, $x$ and $y$, is described by the mean values and widths of
circular uniform distributions. For the velocity coordinates, Gaussian distributions characterized
by their mean values and full widths at half maximum are used. Also the initial time spread of the
molecular beam, which corresponds to the opening time of the valve, is described by a Gaussian
distribution centered around $t_0$. From the initial phase space distribution, the position of a
molecule in phase space is randomly chosen. Then, it is propagated through the beamline, which
includes all mechanical apertures of the experimental setup. From the electric field in the
electrostatic deflector, which is calculated in two dimensions using finite element methods (Comsol
Multiphysics 3.4), and the Stark energy (see Appendix~\ref{appendix:stark}), the force acting on the
molecule in the transverse directions is obtained: $\vec{F} = -\vec{\nabla}W_{\text{Stark}}(E)$. The
electric field is taken to be constant along $z$. Finally, the deflected molecules are propagated
through field-free space to the detection region. For each quantum state, 10$^5$ trajectories are
calculated yielding single-quantum-state deflection profiles $I_s(y)$.

\subsection{Deflection Profiles}
\label{appendix:profiles}

The spatial deflection profile for an ensemble of molecules at a given rotational temperature,
$I(y,T_{\text{rot}})$, is calculated from the single-quantum-state deflection profiles, $I_s(y)$, as
follows:
\begin{equation}
I(y,T_{\text{rot}}) = \frac{1}{w} \sum^N_{s=1} w_s(T_{\text{rot}}) I_s(y)
\end{equation}
Here, $N$ is the number of quantum states included in the simulations and $w_s(T_{\text{rot}})$ is
the population weight for a given quantum state:
\begin{equation}
w_s(T_{\text{rot}}) = g_M g_{\text{ns}} e^{\frac{W_0-W_s}{kT_{\text{rot}}}}
\end{equation}
with $W_0$ being the field-free potential energy of the ground state and $W_s$ the field-free energy
of the current state; $g_M = 1$ for $M=0$ and $g_M=2$ otherwise; $g_{\text{ns}}$ accounts for the
nuclear spin statistical weight of the current state. For both, benzonitrile and iodobenzene,
$g_{\text{ns}} = 5$ for $K_a$ even and $g_{\text{ns}} = 3$ otherwise. The normalization is given by
$w = \sum^N_{s=1} w_s$.

In some cases, a molecular beam cannot be accurately described by a single rotational
temperature~\cite{Wu:JCP91:5278}. Therefore, a two-temperature model is implemented in our
simulations. In the two-temperature model, the low-temperature part of the molecular beam is
calculated as described above. In order to realistically simulate the high-temperature component in
the molecular beam, a very large number of quantum states would have to be included in the
simulations for large asymmetric top molecules. Because these Monte Carlo simulations are very time
consuming, the high-temperature component is approximated by adding a small fraction of undeflected
molecules to the deflected beam profile. This approximation is well justified, because typical
rotational temperatures for the high-temperature component in a molecular beam are on the order of
10~K and, for these temperatures, most of the molecules reside in high rotational quantum states
that have a small Stark shift and, therefore, remain almost undeflected. Thus, the deflection
profile, $I_{\text{2T}}(y,T_{\text{rot}},q)$, in the two-temperature model is calculated as
\begin{equation}
   I_{\text{2T}}(y,T_{\text{rot}},q) = q \cdot I(y,T_{\text{rot}}) + (1-q) \cdot I_{\text{ud}}(y)
\end{equation}
where $0 < q < 1$ and $ I_{\text{ud}}(y)$ denotes the undeflected spatial beam profile that is
obtained when both electrodes are grounded.

\subsection{Fit of Rotational Temperature}
\label{appendix:fit}

In order to calculate the rotational temperature of the molecular beam, the simulated deflection
profiles are fitted to the experimental data using a Nelder-Mead simplex algorithm. In the fitting
procedure, the difference between simulated and measured deflection profiles is minimized, where the
rotational temperature $T_{\text{rot}}$ of the low-temperature component, the fraction $q$ of
molecules in this low-temperature component, and a general intensity scaling factor $s$ of the deflected profiles (with respect to the undeflected beam profile) are used as fitting parameters. All deflection curves
measured for different high voltages are fitted simultaneously to determine the rotational
temperature.

\begin{acknowledgments}
   We thank Henrik Haak for expert technical support. This work is further supported by the
   Carlsberg Foundation, the Lundbeck Foundation, the Danish Natural Science Research Council, and
   the Deutsche Forschungsgemeinschaft within the priority program 1116.
\end{acknowledgments}

\bibliographystyle{jk-apsrev}
\bibliography{string,mp}

\end{document}